\documentclass[12pt]{article}  

\usepackage{graphicx}   

\begin{document}

\title{The mystery of lost energy in ideal capacitors}

\author
{A. P. James$^{1}$\\
\\
\normalsize{$^{1}$Queensland Micro-nanotechnology Center, Griffith University}\\
\normalsize{Nathan, QLD 4111, Australia}\\
\normalsize{E-mail:  a.james@griffith.edu.au}
}

\date{}



\baselineskip24pt

\maketitle

\begin{abstract}
The classical \textit{two-capacitor problem} shows a mysterious lose of energy even under lossless  conditions and questions the basic understanding of energy relation in a capacitor. Here, we present a solution to the  classical  \textit{two-capacitor problem}. We find that by reinterpreting the  energy calculations we achieve no lose of energy  thereby obeying the conservation of energy law.
\end{abstract}

The introductory books in electronic circuits and physics \cite{book1,book2,book3,book4}, often put forward an energy paradox on idealised capacitor switching. This paradox \cite{paper2} is described in Fig. 1a, where energy before and after the switch become closed does not seem to be same.   The main issue here is  the mysterious loss of  50\% energy, despite all the components being ideal (i.e. ideal capacitor, ideal wires, and ideal switches). Further, in Fig.~1a, the total charge, total voltage, and total power in the circuit  is conserved, so having energy reduced by half questions the primary idea of conservation principle. 
For the past several decades, as this paradox had \textit{no explanation} or \textit{solution} under ``idealistic" conditions, much focus has been on rationalising the lose of energy by using non-ideal or realistic conditions such as by the addition of resistors and inductors in the capacitor circuits \cite{paper1,paper2,paper3,paper4,paper5,paper6,paper7,paper8}.


We start with redefining the basic energy-voltage relation for the capacitor circuit as:
\begin{equation}
E_{total}=\frac{1}{2}Q_{total}V_{total}
\end{equation}
where, $E_{total}$ is the total energy in the circuit, $Q_{total}$ is the total charge   in the circuit and $V_{total}$ is the total  voltage (provided by the source)  in the circuit. We also reexpress the relation of charge and voltage  for any $N$ number of capacitors with equal capacitance $C$ as: 
\begin{equation}
Q_{total}=NC|V_c|
\end{equation}
 where $|V_c|$ is the magnitude of the voltage across each capacitor with capacitance $C$. This idea can be expressed using a water-charge analogy (Fig. 1b), the question in general is to distribute the volume of water from a main tank (the source) equally among storage tanks with same volume and dimensions.
Here charge in a capacitor corresponds to volume of water, voltage corresponds to the level (height) of water and capacitance is related to the capacity of the tank.

Using this redefined view on the energy-voltage relation, we  attempt to solve the \textit{two-capacitor problem}. The equivalent circuit for this problem can be drawn as shown in Fig. 1c. In analogy to the water-charge model in Fig 1b, irrespective of how the capacitors are tied up, the total charge (or volume of water) before and after should remain same, in other words, the volume of water pumped from the main storage tank should be equal to sum of volume of water  received at the storage tanks. In the two-capacitor problem, we can think this analogy as the following: (1) the water from the main tank is pumped to a first tank till it reaches a specified height (voltage =$V$), and (2) the main supply is removed and the water from the first tank is now pumped to a second tank until both tanks have  equal volume of water. Here, it maybe noted that we do not take any non-ideal  assumptions on  the storage tank mechanisms and keep all components models in the circuit as ideal. The leftmost circuit in Fig. 1c shows the equivalent circuit for charge storage mechanism that occur as a result of pumping the charge from a main tank (source) to a storage tank with capacitance $C$. The storage tank is analogous to a capacitor with capacitance $C$ and main tank analogous to the source voltage with voltage $V_{total}=V$, we can calculate the  initial  total charge $Q_{initial}=CV=Q_{total}$, initial total  voltage $V_{initial}=V=V_{total}$ and  initial  total energy $E_{inital}$   as:

\begin{equation}
E_{inital}=\frac{1}{2}Q_{inital}V_{inital}=\frac{1}{2}CV^2
\end{equation}

When the switch is closed (Fig. 1a) at time $t=0^+$, both capacitors (with equal capacitances $C$) become connected, and the charged capacitor now charges the newly connected capacitor until the voltage and charge across each capacitor at equilibrium reaches a value of $V/2$ and $Q_{initial}/2$.  According to the principle of conservation of charge, and from the water-charge analogy, irrespective of how the capacitors are connected the total charge before and after the switch is closed  should remain the same. However, we find that direct application of $Q=CV$ relation contradicts this very basic assumption. This can be understood better through two examples of series and parallel combination of capacitors. As a general case for a series combination of capacitors, there are two possible ways by which charge  reflects on the individual capacitors having a capacitance $C$ with potential difference $\frac{V}{2}$. Depending on polarity of the charge, the voltage across the equivalent capacitance $C/2$ can be $\frac{V}{2}+\frac{V}{2}=V$ or $\frac{V}{2}-\frac{V}{2}=0$, thereby reducing the total  charge to $Q/2$ or 0 respectively. The equivalent circuit models in Fig 2a illustrates these two situations.
 In analogy to the water-charge model, two individual storage  tanks are placed one above the other (Fig. 2a). Through the top view, the level information of the water (charge) is not visible and is lost ($\frac{V}{2}-\frac{V}{2}=0$) , while from the side-view individual level information of the water (charge) is visible and so added ($\frac{V}{2}+\frac{V}{2}=V$). 
The equivalent capacitance of the storage tank is proportional to the ratio between total area and total height. Irrespective of whether its a side-view or top-view,  since in Fig. 2a the total area remain constant while total height of the combined tank doubles, the overall capacitance decreases by a factor of two. Now, although the equivalent capacitance seems to reduce the total capacitance by a factor of two, the capacity of the individual storage tanks do not change and so stores the same amount of charge.  It can be further observed that in the calculation of charge, the use of top view capacitor model can be straight away avoided as it loses the information on charge levels (voltage $=0$) and causes the total charge to reflect as zero.  On the other hand, if we consider the side view model where the total voltage is $V$ and the equivalent capacitance  $C/2$. When calculating the charge, the reduction in capacitance reduces the total charge  by a factor of two, which is physically incorrect and again contradicts the total amount of already  available charge in the tank. So the only possible solution in both cases (side view and top view) is to calculate the total charge as a sum of charges contributed by individual tanks, which if using $Q=CV$ relation for each capacitor with capacitance $C$ would mean to always take the magnitude of individual voltage (as $Q=C|V|$).

Figure 2b shows the situation when the capacitors are connected in parallel. The tanks are now placed one beside the other. Irrespective of whether its a side-view or top-view, the  total area of this combination  increases by a factor of two and therefore the equivalent capacitance increases by a factor of two. In the side view, the voltage level does not change, and remain at $Vc=V/2$.     In the case of the  circuit models illustrated for  side view (Fig. 2b), putting a probe across the capacitor would yield a value of $V/2$,  its equivalent capacitance  will be $2C$ and charge $Q=CV$. Here, the charge relation work fine as the capacity of the tanks become additive and the levels of voltages are not disturbed.
However, the top view of the water-charge model (Fig. 2b)
result in loss of depth information, $\frac{V}{2}-\frac{V}{2}=0$ and cannot be used for charge calculations. Here, again the only possible solution  is to account for charge in  each storage tanks individually by considering the magnitude of individual tank voltages for the charge calculations.
Clearly, the idea of equivalent capacitances to calculate the total charge do not comply properly with the physical meaning of charge storage. Individual treatment of capacitor is needed for preserving the physical meaning of charge for energy calculations. Since the polarity of  voltage is not important for charge calculation, it should not be important for energy calculations as well (which is in agreement with the $\frac{CV^2}{2}$ relation). 

Using these ideas, we revisit the  two capacitor problem when the two capacitors  are connected as shown in Fig. 1c.  Each capacitors have a capacitance of $C$ and voltage of $V/2$ across it. We can calculate the final total charge $Q_{final}=C|\frac{V}{2}|+C|\frac{V}{2}|=CV=Q_{total}$ and show conservation of charge as, $Q_{final}=Q_{initial}$.
We can  calculate the final total voltage provided by the voltage source $V_{final}=|\frac{V}{2}|+|\frac{V}{2}|=V=V_{total}$ as the magnitude sum of potential difference that occur across the individual charge tanks, this results in the conservation of voltage as, $V_{final}=V_{initial}$. 

Substituting the values of $Q_{final}$ and $V_{final}$ in Eq. (1), we calculate the final energy $E_{final}$ after the switch is closed as: \begin{equation}
E_{final}=\frac{1}{2}Q_{final}V_{final}=\frac{1}{2}CV^2
\end{equation}
From (3) and (4), it can be seen that energy is also conserved as, $ E_{final}=E_{initial}$. 

In this letter, we present a simple and general solution to the energy conservation paradox shown by classical \textit{two-capacitor problem}. As shown by Fig. 2, although the total charge should remain the same in all the possible capacitor combinations, but application of conventional $CV$ relation using equivalent circuit models makes conservation of charge  fail, henceforth energy is also not conserved. However, such a situation is physically impossible in ``idealistic" or ``realistic" conditions. When the  physical meaning of the charge and voltage is preserved, by applying the conservation of voltage and charge separately, we result in an energy equation that is general to use in any type of  capacitor based circuit analysis. By this approach we are able to completely resolve the paradox that exists in the  ``idealistic"  conditions.

%

\clearpage

\begin{figure}
	\includegraphics[width=120mm]{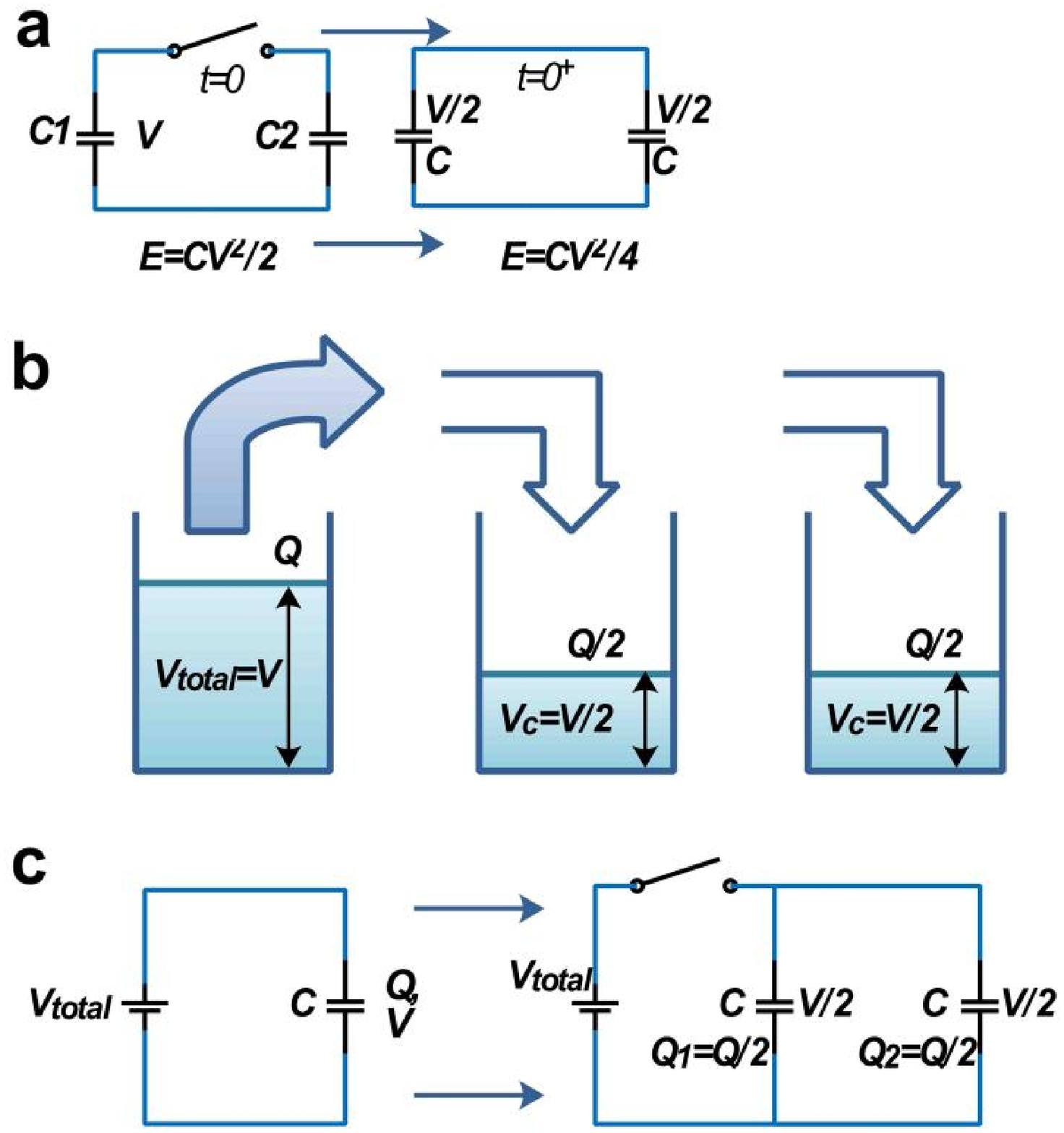}
	\caption{The circuit diagrams for illustrating the two-capacitor problem. {\bf(a)} shows the circuit diagram for the problem, switch is open initially at time $t=0$ and closed soon after at time $t=0^+$,  {\bf(b)} analogy of water-storage tank to a capacitor charge tank. The water from a main storage tank is equally pumped to two storage tank. The main storage tank is analogous to a voltage source, and {\bf(c)} an equivalent circuit using the capacitor charge tank model.}
	\label{fig:1}
\end{figure}

\clearpage

\begin{figure}
	\includegraphics[width=120mm]{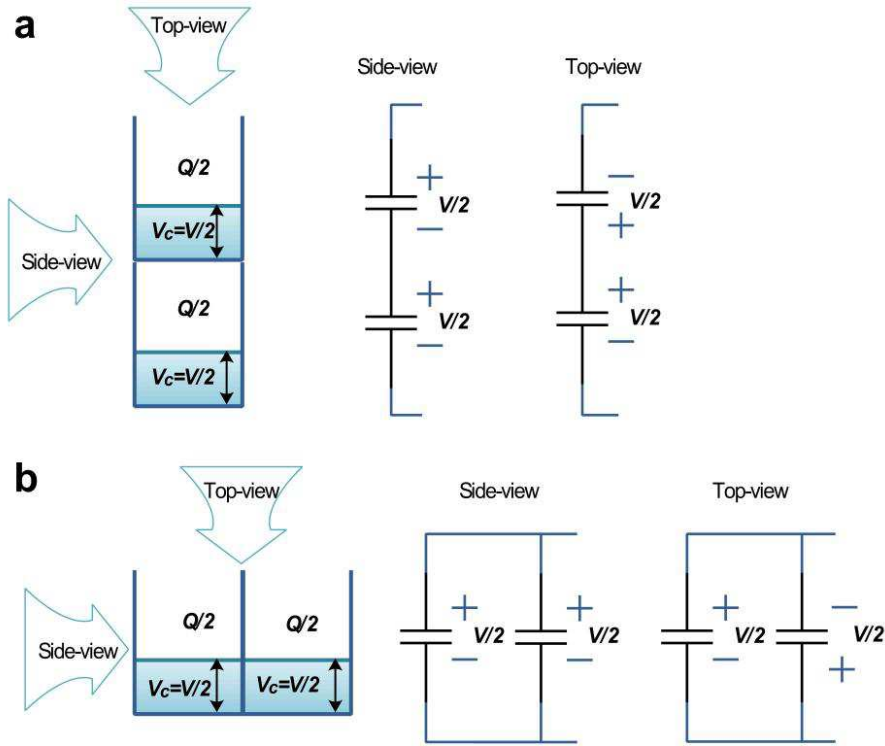}
	\caption{Illustration shows the equivalent capacitor models using the proposed water charge  analogy. {\bf(a)} shows the analogy of series capacitor configurations. Side view preserves the  depth information and the individual voltage level contributions are added in the process. Top view results in total loss of  depth information and the individual voltage level contributions are lost in the process, and {\bf(b)} shows the analogy of parallel capacitor configurations.
Side view preserves the  depth information but the individual voltage level contributions are not added and partly lost in the process.
Top view results in total loss of  depth information and the individual voltage level contributions are lost in the process. }
	\label{fig:2}
\end{figure}

\end{document}